%
%


\def\msun{{\rm\,M_\odot}}

\def\K{{\rm\,K}}

\def\\{\hfil\break}
\def\newpage{\vfill\eject}
\def\lsim{\raise0.3ex\hbox{$<$}\kern-0.75em{\lower0.65ex\hbox{$\sim$}}}
\def\gsim{\raise0.3ex\hbox{$>$}\kern-0.75em{\lower0.65ex\hbox{$\sim$}}}
\def\etal{{\it et al.~}}

\magnification=\magstephalf
\hoffset = -.22 truein
\voffset = -.48 truein
\hsize=18.0truecm
\vsize=25.0truecm
\pageno=1
\parskip=11pt
\parindent=0pt
\baselineskip=11pt

\font\titlefnt = cmbx12 scaled \magstep 1


\ 
\\
\\
\\
\\
{\titlefnt HYDRODYNAMICS OF ACCRETION ONTO BLACK HOLES}
\\
\\
\\
M.-G. Park${}^{1,2}$ and J. P. Ostriker${}^1$
\\
\\
\\
{\it ${}^1$ Princeton University Observatory, Princeton, NJ 08544, USA}
\\
{\it ${}^2$ Kyungpook National University, Department of Astronomy and
Atmospheric Sciences, Taegu 702-701, KOREA}
\\
\\
\\
\\
ABSTRACT
\\
\\
Spherical and axisymmetric accretion onto black holes is
discussed. Physical processes in various families of
solutions are explained and their characteristics are
summarized. Recently discovered solutions
of axisymmetric flow provide us with various 
radiation efficiency and spectrum, which may successfully
model diverse accretion systems. Possible role of
preheating is also speculated.
The various families of solutions can be plotted
as trajectories on the plane of $(l,e)$
[or $(l,\dot m)$] where $l$ is the luminosity
in units of the Eddington luminosity, $\dot m$ is
the similarly defined mass accretion rate and
$e$ is the efficiency defined by 
$e \equiv L/\dot M c^2 = l/\dot m$. 
We discuss the domains on these planes where
solutions are known or expected to be unstable
to either spherical or non-spherical perturbations.
Preliminary analysis indicates that a preheating instability
will occur along the polar direction of the advection
dominated flow for $e ~\gsim~ 10^{-2}$.
\par
I. INTRODUCTION

The generic accretion flow can be broadly classified as
(quasi) spherical or axisymmetric one. The angular
momentum of the flow and its transfer processes
will determine
whether the flow is spherical or not. True spherically
symmetric flow is possible
only when the angular momentum is rigorously zero. 
However, when the angular
momentum is small enough and there exists 
a process that can effectively 
remove or transport the angular momentum, 
then the flow maintains
a roughly spherical shape (Loeb and Laor 1992; Narayan and Yi 1995a). 
If the angular momentum is significant, the flow becomes 
rotationally supported and is flattened to disk.

Until recently, most works on spherical accretion flow have focused
on zero angular momentum flow and those on axisymmetric flow on
thin flat disk. Purely spherical flow is physically
well defined and has high degrees of symmetry, 
which make it possible to treat 
hydrodynamics, radiative transfer and various gas-radiation microphysics 
rather accurately. Yet it
might be too ideal for real astrophysical situations.
The disk flow produces more radiation for a given mass accretion
and the solution is well understood in some parameter ranges
except the angular momentum transport process remains uncertain. 
However, the true solution of the disk flow
requires solving the two-dimensional hydrodynamic and radiation
transport equations, a quite daunting challenge even with
today's enormous computing power.

In this review, we use the dimensionless luminosity
$ l \equiv L / L_E $
and the dimensionless mass accretion rate
$ \dot m \equiv \dot M / \dot M_E $
where $L$ is the luminosity of the accretion flow,
$L_E$ the Eddington luminosity,
$\dot M$ the mass accretion rate and
$\dot M_E$ the Eddington mass accretion rate,
$ \dot M_E \equiv L_E/c^2 $.
In some literature, the Eddington accretion rate divided by the
radiation efficiency 
$ e \equiv L / (\dot M c^2) = l/{\dot m} $
is used as the basic unit. 
In accretion onto a black hole,
$e$ can be arbitrary unlike in accretion onto a neutron star.

II. SPHERICAL FLOW

The first study of spherical accretion onto 
compact objects dates back more than forty years (Bondi 1952).
In this classic work, hydrodynamics of polytropic flow is
studied within the Newtonian dynamics, 
and it is found that either a settling or
transonic solution is mathematically possible 
for the gas accreting onto compact objects.
Especially, the accretion rate is highest 
for the transonic solution.
Relativistic version of the same problem was solved
by Michel (1972) twenty years later,
after the discovery of celestial X-ray sources.
He showed that accretion onto the black hole should
be transonic. 

However, the real accretion necessarily involves 
radiative processes, which make the flow non-adiabatic and,
at the same time, more diverse and interesting. 
In general, compression of the infalling gas raises
the temperature of the accreting gas and then
radiation is
produced by the radiative cooling of heated gas,
which in turn interacts with the other part of the flow
via scattering and absorption,
thereby heats or cools the upstream gas. 
So understanding the accretion
flow involves simultaneously solving for the gas and radiation
field for given outer boundary conditions. 
This is particularly important for accretion onto black holes
because the lack of the hard surface in a black hole
makes the radiation efficiency unknown a priori. 
In accretion onto neutron star, 
the surface potential of the star determines how much luminosity
is generated for a given accretion rate, regardless of 
the details in radiative and hydrodynamic processes.
In accretion onto black hole,
the radiation efficiency is determined by the state of the gas at
all radii. If the gas is kept at low temperature, most of the
gravitational potential energy would be lost to the hole in the form
of gas kinetic energy. If the gas is kept at high temperature
by the radiative heating or other processes, the luminosity 
is increased and less kinetic energy is lost to the hole.

Spherical accretion onto a black hole is generally specified by
the mass accretion rate and by
the luminosity and the spectrum of the radiation field.
As long as two body processes are important, 
the solutions become 
scale free and are applicable
to black holes of arbitrary mass (Chang and Ostriker 1985).
These scale-free solutions then depend primarily on 
the dimensionless parameters 
$l$ and $\dot m$ or $l$ and $e$ (Ostriker \etal 1976).
However, if we require the self-consistency between the gas
and the radiation field, only certain combinations of
the accretion rate and the luminosity (plus spectrum) are
allowed, that is a given type of solution is described
by a line on the $(l,\dot m)$ plane.
Now we describe
these solutions one by one in the order of increasing $\dot m$.

\vskip 4.6truein

\leftskip=1truein 
\rightskip=1truein
Fig. 1. Spherical and axisymmetric accretion flow solutions in 
        luminosity vs mass accretion rate plane. 

\newpage
\ \vskip 4.6truein

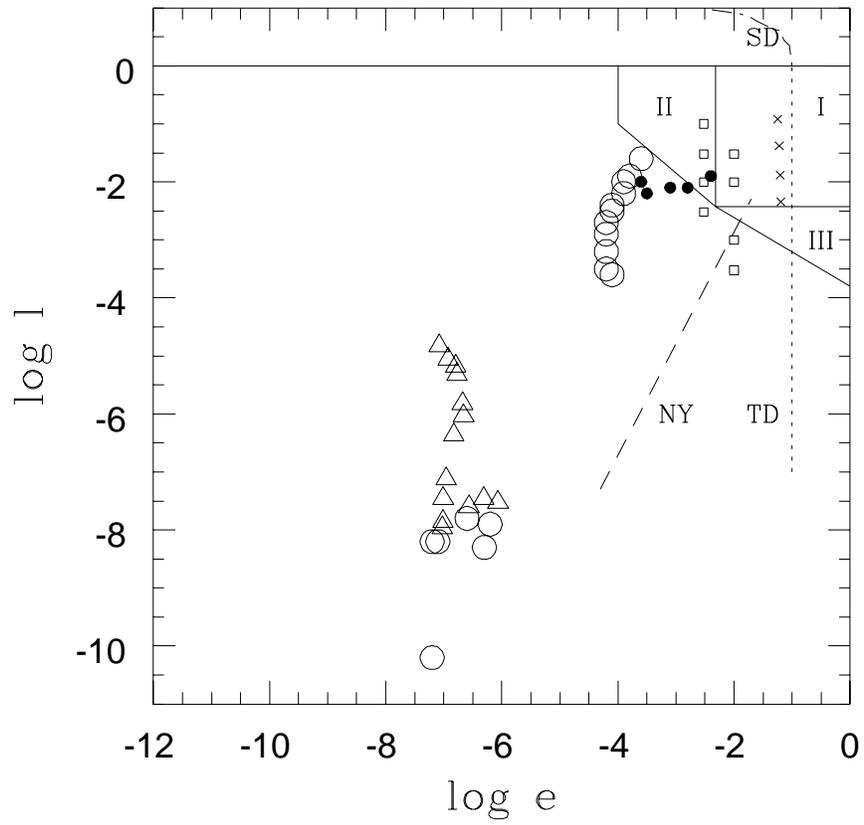
Fig. 2. Spherical and axisymmetric accretion flow solutions in
        luminosity vs radiation efficiency plane. 

\leftskip=0in
\rightskip=0in

\underbar{2.1 Solutions with $\dot m \ll 1$}

Since the electron scattering optical depth from infinity
down to the horizon is roughly given by 
$ \tau_{es} = {\dot m} (r / r_s)^{-1/2} $
where $r_s$ is the Schwarzschild radius,
$\dot m \ll 1$ flow is optically thin to scattering,
and whatever photons produced escape without difficulty.
Any kind of interaction between 
photons and gas particles is minimal.
Gas is freely falling so $v \propto (r/r_s)^{-1/2}$ and 
its temperature has the virial value, 
$T \propto (r/r_s)^{-1}$,
until electrons become relativistic (Shapiro 1973).
In Newtonian approximation, 
$\gamma = 5/3$ accreting flow has a constant Mach number,
because the sound speed is increasing as fast as 
the infall velocity (Bondi 1952). \\

But in the relativistic regime, 
the sound speed or the electron temperature is 
increasing less rapidly and the flow becomes
supersonic with increasing Mach number 
for $T ~\gsim~ 10^9 \K$ (Shapiro 1973, Park 1990a).
The temperature can reach as high as few $10^{10}\K$,
producing a hard relativistic bremsstrahlung spectrum.
Yet the luminosity is quite small due to the very low density.
Any radiative heating, especially Compton heating,
for this family of solutions can be ignored
because of the low luminosity.
Typical solutions are marked as {\sl large empty
circles} in $\dot m < 1$ region of Figures 1 and 2
and their radiation efficiency $e$ is less than $10^{-8}$ (Park 1990a).

\underbar{2.2 Low-Temperature Solutions with $\dot m \gsim 0.1$} 

As $\dot m$ approaches 0.1, bremsstrahlung and atomic cooling
becomes more efficient and 
gas cools down  $\sim 10^4\K$, the
equilibrium temperature of the surrounding gas. 
Depending on the mass of the hole, the flow
can form an effectively optically thick${}^1$
\footnote{}
{~~~${}^1$~$\tau^\ast \equiv (3\tau_{es}\tau_{abs})^{1/2} \gg 1$
where $\tau_{abs}$ is the absorption optical depth. }
core with blackbody radiation
field inside. This regime represents the transition
from adiabatic to non-adiabatic flow (Park 1990a; Nobili \etal 1991).

As the gas density at the boundary increases further, so does the
mass accretion rate, and the flow becomes effectively optically thick
out to a larger radius. The flow is maintained at a low temperature.
These solutions are quite similar to those in a stellar 
interior except that the flow is infalling
by the gravity and $PdV$ work is the source of the luminosity 
(Flammang 1982, 1984; Soffel 1982; Blondin 1986; Nobili \etal 1991).
Typical solutions are shown in Figures 1 and 2 as {\sl large triangles}
(Nobili \etal 1991).

An important and interesting radiation transport process called
radiation trapping happens inside certain radius
where $ v > c/(\tau_{es}+\tau_{abs}) $ is satisfied.
When the diffusion speed of photons is slower than the flow's
bulk velocity, most of the photons 
are carried inward with the flow (Begelman 1978).
The correct treatment of this process requires relativistic
radiative transport equations, 
at least up to the order $(v/c)^1$
(Flammang 1982, 1984; Park 1990a; Nobili \etal 1991; Park 1993).
In relativistic flow, especially in an optically thick one, 
the flux seen by the observer comoving with the flow and the flux
seen by the stationary observer (relative to the coordinate)
should be carefully distinguished.
The momentum transferred to the gas from the radiation is closely
linked to the former, a comoving-frame flux, 
and the luminosity seen by the observer
at infinity to the latter, a fixed-frame flux. 
(Mihalas and Mihalas 1984; Park 1993).
For an ccretion flow, the comoving flux is always larger than the 
fixed frame flux and should be less than the Eddington flux 
for steady inflow (Park and Miller 1991).
This radiation trapping and the low temperature of the flow make
this family of solutions very inefficient radiators with 
$e \sim 10^{-7}$.

\underbar{2.3 High-Temperature Solutions with $\dot m \gsim 0.1$} 

All this would change if gas can be preheated by outcoming
hard radiation produced at smaller radii.
The existence of higher temperature, higher luminosity
accretion solution due to this preheating was first proved
by Wandel \etal (1984) in simplified treatment. A more accurate
and relativistic treatment show that these higher luminosity
and higher efficiency steady-state solutions exist only for
$ 3 ~\lsim~ \dot m ~\lsim~ 100$ 
(Park and Ostriker 1989; Park 1990a,b; Nobili \etal 1991).
So there exist two different families of solutions for given
$\dot m$, low temperature and high-temperature one (Fig. 1).

The gas temperature can reach up to $10^9 \sim 10^{10} \K$;
the flow is optically thick to scattering, and the radiation 
trapping occurs near the hole. However, it is optically thin
to absorption, $\tau^\ast \ll 1$. Bremsstrahlung
photons produced in the inner region are upscattered 
by hot electrons around, which subsequently
heat cool electrons in the outer part
by Comptonization. These solutions are plotted in Figures 1 and 2
as {\sl large empty circles} (Park 1990a; Nobili \etal 1991). 
These solutions are much more luminous, 
$l ~\simeq~ 10^{-4}-10^{-2}$, and produce much harder photons
than the low-temperature
solutions of the same mass accretion rate.
Still, they are much less efficient radiators, $e \sim 10^{-4}$,
than the thin disk.

There are ways to increase the radiation efficiencies.
The dissipational heating of turbulent motion and the
magnetic field reconnection (M\'esz\'aros 1975;
Maraschi \etal 1982) and self-consistently sustained
high electron-positron pair production 
(Park and Ostriker 1989) are two examples.
Solutions of the former type are shown as {\sl small
squares} (Maraschi \etal 1982) with $e$ as high as $\sim 0.1$
and the latter type
as {\sl small filled circles} (Park and Ostriker 1989) in Figures
1 and 2. They constitute yet another families of solutions
for given $\dot m$. 
Though these solutions are quite attractive because of the
high efficiency and high temperature, they are quite likely
to be subject to the preheating instability described below. 
The reason that these solutions are found in steady-state
calculation at all is that
either preheating is ignored (M\'esz\'aros 1975; Maraschi \etal 1982)
or only the inner part of the flow is considered 
(Park and Ostriker 1989).
It is possible to maintain the steady flow under significant
preheating by having a shock (Chang and Ostriker 1984).
However, no self-consistent solutions with a steady shock is
yet constructed, although there is some indication that such solution
might exist around $\dot m \sim 1$ or $\dot m \sim 100$ 
(Park 1990b; Nobili \etal 1991).

\underbar{2.5 Preheating} 

When radiation and gas interact, both momentum and
energy are transferred from one to the other. The luminosity
of a given spherical accretion flow has a well
known upper limit above which photons would 
give the gas particles too much momentum to 
infall steadily. When gas is fully ionised and
opacity is dominated by the Thomson
scattering, this limit is the well-known Eddington luminosity.

Ostriker \etal (1976) found another limit 
based on the energy transfer. They found that if the gas
is preheated too much around the sonic point, it gets too
hot to accrete in steady-state. When the gas temperature at the
sonic point is $10^4 \K$ and the Compton temperature of the
preheating radiation is $10^8 \K$, the disruption of the accretion
can occur at $l$ less than 0.01.
This is why Park (1990a,b) and Nobili \etal (1991)
could not find high-temperature 
self-consistent solutions for $\dot m ~\lsim~ 3$
and $\dot m ~\gsim~ 100$: in the former parameter region,
the photon energy of the preheating radiation is too high
and in the latter, the luminosity is too high.

This disruption of flow by preheating happens
in a square region labeled I in Figures 1 and 2. 
The exact location
of the boundary depends on various conditions of the gas
and the radiation field as well as the shape of the cooling
curve. Under certain conditions, the region could be much
smaller than the one shown in Figures 1 and 2 (Ostriker \etal
1976; Cowie \etal 1978; Bisnovatyi-Kogan and Blinnikov 1980;
Stellingwerf and Buff 1982;  Stellingwerf 1982; Krolik and
London 1983). 

The time-dependent behavior of the preheated flow was investigated
by Cowie \etal (1978) with analytic analysis and hydrodynamic
simulations (see also Grindlay 1978). 
They found two distinct type of time dependent 
behaviors both inside and near the boundaries of region I.
High luminosity, low efficiency flow (region II in
Figs. 1 and 2) develops
preheating inside the sonic point, 
which produces recurrent
flaring in short time intervals,
$\sim 100(M/\msun)~\hbox{sec}$, with its average
luminosity similar to the steady-state solutions.
On the other hand, in higher efficiency, lower luminosity flow 
(region III in Figs. 1 and 2) preheating outside the sonic
point induces longer time scale,
$\sim 10^9(M/\msun)~\hbox{sec}$, changes in accretion rate
and luminosity.
Even the solution outside the preheating
regime appears to have short time variability due to
the preheating (Zampieri \etal 1996).

The fact that high luminosity, high efficiency 
steady-state solutions would
suffer the preheating instability implies the
time-dependent nature of these solutions. 
As the luminosity and efficiency of the accretion increase,
we expect more variability in the flow and in the 
emitted radiation in general.

III. AXISYMMETRIC ACCRETION

\underbar{3.1 Thin Disk}

Until recently most works on non-spherical accretion have
focused on the thin disk accretion
(see Pringle 1981, Treves \etal 1988, 
Chkrabarti 1996 for reviews). Since the pioneering work by 
Shakura and Sunyaev (1973), Pringle and Rees (1972),
and Novikov and Thorne (1973), 
thin disk models with so-called $\alpha$ viscosity
have been successfully applied to many astronomical
objects including various X-ray sources 
(see Frank \etal 1985 for reviews). 

Thin disk accretion model has merits of being simple: 
the equations become one-dimensional (in radius) and all physical
interaction can be described by local quantities.
Gas is rotating with Keplerian angular momentum, 
which is transported radially by viscous stress.
The gravitational potential energy is locally converted to heat
by this viscous process. The gas cools by radiating
in the vertical direction, thereby not affecting any other
part of the flow. The radiation efficiency depends only
on the location of the inner disk
boundary and is generally $\sim 0.1$.
So any thin accretion
disk model lies on the {\sl dotted line}
$e = 0.1$, denoted as TD, in Figures 1 and 2. 
However, its application cannot
be extended to high luminosity systems due to the instability
at a high mass accretion rate (Lightman and Eardley 1974;
Shakura and Sunyaev 1976; Pringle 1976; Piran 1978)
nor to hard X-ray sources due to the
low temperature of the disk. Although Shapiro \etal (1976)
discovered another family of thin disk solutions that have very
high electron temperature, they are also found to be thermally
unstable (Pringle 1976; Piran 1978). 

\underbar{3.2 Slim Disk}

If the dimensionless mass accretion rate $\dot m$ approaches 
$ e^{-1} $, the vertical height of the disk becomes comparable
to the radius and the disk is not thin any more.
In this regime, 
the radial motion of the gas becomes important and 
the angular momentum of the gas is below the Keplerian
value unlike in thin disk.
Abramowicz \etal (1988) improved over the thin disk approach by
incorporating the effect of pressure in the radial motion
to find new type of solutions in high $\dot m$ regime.
They are stable against the viscous and thermal instabilities.
Viscously dissipated energy can now be
advected into the hole along with the gas in addition to being
radiated away through the surface of the disk. They named
these solutions `slim disk' because the disk is not
thin, yet not so thick that vertically integrated 
equations are valid. This work shows
that disk accretion too can have efficiency
other than $\sim 0.1$.
These solutions are represented in Figures 1 and 2 
as the {\sl dot-dashed curve} (SD) at the
end of $e = 0.1$ line (Szuszkiewicz \etal 1996).

\underbar{3.3 Self-Similar Flow}

To get the slim disk solutions,
Abramowicz \etal (1988) had to explicitly integrate
the hydrodynamic equations that have a critical point.
However, recently 
Narayan and Yi (1994) found that the equations admit
self-similar solutions if the advective cooling
is always a constant fraction
of the viscous dissipation. These solutions are
very simple and at the same time very restrictive.
The density, velocity, angular velocity, and the total
pressure are simple power laws in radius:
$\rho \propto r^{-3/2}$, $v \propto r^{-1/2}$,
$\Omega \propto r^{-3/2}$, and $P \propto r^{-5/2}$. 
Unfortunately the temperature
of these solutions is always close to the virial
value and the vertically integrated equations
may not be valid.

They addressed this question in the next work (Narayan
and Yi 1995a). By assuming the self-similar form of physical
quantities in radius, the axisymmetric two-dimensional hydrodynamic
equations are reduced into one-dimensional equations 
in polar angle only, and
easily solved. The solutions show a radial velocity
of a few percent of free-fall value
at the equatorial plane and zero radial velocity 
on the pole. So the gas is preferentially accreted 
along the equatorial plane. However, the flow is always
subsonic because the pressure is close to the
virial value. Since only the transonic accretion is allowed,
this makes the solution inapplicable close to the hole${}^2$.
\footnote{}{~~~${}^2$~
However, recent global solutions of slim disk equations
show that self-similar solutions describe the flow
reasonably well at intermediate radii (Narayan \etal 1996b;
Chen \etal 1996).}

One important result of this work is that
physical quantities calculated from the 
vertically integrated equations generally agree 
with the polar angle averaged quantities. This justifies
using vertically integrated 
one-dimensional equations, i.e., slim disk equations, 
even to the two-dimensional accretion flow.
But rigorously, 
this convenience applies only to the self similar
solutions.

Due to the simplicity of the solutions, many diverse
microphysics can be incorporated into these solutions
without much difficulty (Narayan and Yi 1995b).
The typical solutions with low $\dot m$ have efficiency
much smaller than the standard thin disk (dashed line in Figs.
1 and 2).   
These advection-dominated 
accretion models with low $\dot m$ successfully
explain various low luminosity sources
which have been hard to model with highly
efficient thin disk,
like Sgr A${}^\ast$ (Narayan \etal 1995), NGC 4258
(Lasota \etal 1996), and X-ray transients 
(Narayan \etal 1996a; Chen and Taam 1996). 
Typical solutions are denoted as NY in Figures
1 and 2 (Narayan and Yi 1995b).
All these models have most of the gravitational
potential energy of the accreted matter advected into the hole
without ever being converted to the radiation.
So these flows are quite similar to the almost
adiabatic, high temperature, low $\dot m$ spherical
solutions of Shapiro (1973). In a way it should be
because the above self-similar forms of various physical
quantities are exactly those in the non-relativistic
adiabatic spherical accretion flow.
In fact, optically thin
two-temperature spherical accretion with a magnetic field also
produces a spectrum that roughly agrees with 
that of Sgr A${}^\ast$ (Melia 1992, 1994).  
Another interesting feature of these models is that
because the total pressure is a constant fraction of the
virial value, the flow has to be two-temperature
and/or some magnetic field should exist, otherwise the radiation
spectrum will be simply that of the free-fall, virial temperature
plasmas. 

\underbar{3.4 Unified Description}


All these seemingly different solutions are actually just 
different families of solutions 
for the axisymmetric accretion flows with angular momentum.
In each family, appropriate assumptions are made to ease the
difficulty of solving the relevant equations.
Specific name is attached to each family
to describe the physical characteristics of the flow,
e.g., thin disk, slim disk, and advection dominated flow.
Although slim disk solutions and self-similar flow solutions
are the results of different approach---the former
from the one-dimensional approach and the latter from the
self-similarity in two-dimensional flow---their final results
agree in general, albeit not in detail, and the description
of these axisymmetric flows in consistent manner
has become finally possible (Chen \etal 1995).

A very revealing way of presenting these solutions is to look
at the relation between the mass accretion rate $\dot m$ 
and the surface mass density $\Sigma$ at a given radius. 
Different families of solutions appear as 
specific curves in $\dot m$ vs $\Sigma$ plane (Fig. 3).
${}^3$\footnote{}
{~~~${}^3$~The solutions are for $\alpha=0.01$,
$M=10\msun$, and $r=20GM/c^2$. The surface mass density
$\Sigma$ is in $\hbox{g/cm}^3$.}
Since the shape of the curves depends on the mass of the
hole and the value of the viscosity parameter $\alpha$,
we describe a specific case where 
$M = 10 \msun$ and $\alpha = 0.01$ (Chen \etal 1995).
Solutions can be broadly separated into (vertically)
effectively optically thin and thick flows. 
The left {\sl dotted curve} (SLE,NY) in Figure 3 
represents the former and
the right {\bf S} shape {\sl solid curve} 
(GTD,RTD,SD) the latter.
The former exists only for $\dot m \ll 1$ and
the latter for any value of $\dot m$.

The lower (GTD) and middle (RTD) branch of {\bf S} curve are the
classic gas pressure dominated and radiation pressure
dominated thin disk solutions of Shakura and Sunyaev (1973).
In both, radiative cooling dominates over the advective
one, and the lower branch is thermally and viscously
stable while the middle branch is unstable to both type
of instability. The upper (SD) branch of {\bf S} curve
is the original slim disk solution in which the advective
cooling dominates and stabilizes the radiation pressure 
dominated flow (Abramowicz \etal 1988). 
This solution is not geometrically thin.
The disk scale height can be comparable to the
cylindrical radius, especially when $\dot m > e^{-1}$. 
The temperature of these {\bf S} curve solution is not high,
$\lsim~ 10^8 \K$, 
and the radiation spectrum
will be the superposition of modified blackbodies
at different temperatures. 
However, they are very efficient
radiators, $e \sim 0.1$. 

\vskip 3.2truein

\leftskip=1truein 
\rightskip=1truein

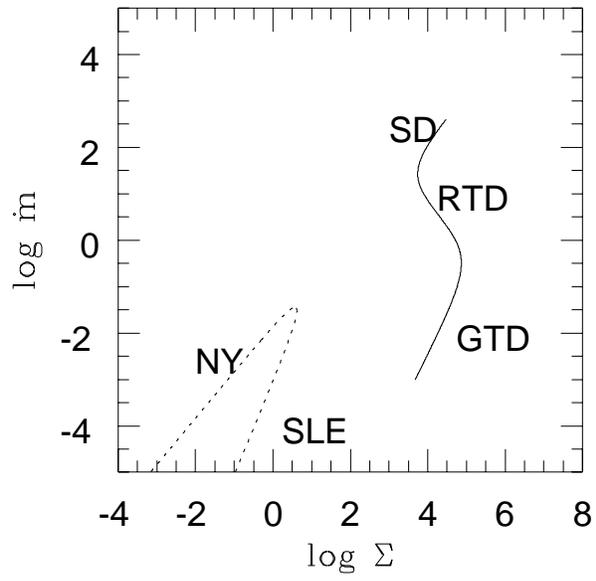
Fig. 3. Various axisymmetric accretion solutions in accretion
        rate vs surface mass density plane. 

\leftskip=0truein 
\rightskip=0truein

The right (SLE) branch of the effectively optically thin
solutions represent the two-temperature hot disk
solutions of Shapiro \etal (1976), which are
viscously stable yet thermally unstable (Pringle 1976;
Piran 1978) and radiative cooling dominated. 
The left (NY) branch is the advection dominated flow
of Narayan and Yi (1994) and Abramowicz \etal (1995), 
in which most of the viscously
generated heat is advected into the hole stabilizing
the flow. The gas temperature is very high,
$\sim 10^9 \K$, and the radiation
spectrum will be that of Comptonized synchrotron 
and bremsstrahlung.
The radiation efficiency of the right branch is always $\sim 0.1$
and that of the left branch has roughly $l \propto \dot m ^2$
and $ e \propto \dot m$ as in optically thin spherical
accretion (Narayan and Yi 1995b). 

As in spherical accretion, there can be other interesting
family of solutions if we consider the pair production 
(Kunsunose and Mineshige 1992 and references therein) 
or various type of shocks
(Chakrabarti 1996 and references therein),
which we will not discuss further.

\underbar{Preheating and Outflows}

One interesting aspect of axisymmetric accretion unexplored so
far is the effect of preheating.
In thin disk flow, any part
of the disk is free from the radiation 
produced in the other parts of the
disk either due to the geometry (vertical direction) or 
due to the high absorption optical depth (radial direction).
But in true two-dimensional flow, hard radiation generated
at the inner part of the flow should go through the outer part.
This radiative heating will be more complex than
in a spherical flow. For example, the self-similar flow
of Narayan and Yi (1995a) has zero radial velocity along
the pole, around which viscous heating, radiative cooling, and
advective cooling are very small compared to 
those in the equatorial region.
So this part of the flow can be very easily heated by Comptonization 
to a high temperature, possibly higher than the virial value.
It is quite plausible that this could develop some 
interesting phenomena, like 
relaxational oscillations seen in spherical flow (Cowie \etal 1976)
or simply outflow. Two-dimensional flow with steady accretion
onto the equatorial plane and time-dependent accretion or 
outflow in the polar direction will be quite 
useful in explaining some high energy
sources. We know that time-dependence and outflow are
the norm rather than the exception in these sources.
Thus the potential instability of the self-similar
solutions to preheating in the polar region should be
seen as a virtue of the solution.
A preliminary anaysis indicates that a preheating instability
will first occur along the polar direction
for $e ~\gsim~ 10^{-2}$.
Two-dimensional numerical hydro simulations show that 
accretion and outflow can coexist (Chakrabarti and Molteni 1993;
Molteni \etal 1994; Ryu \etal 1995; Molteni \etal 1996). 
Preheating might be yet another ingredient 
in the outflow and the time variabilities in accreting flows.

IV. CONCLUSION

Rapid and exciting recent developments in the theory of accretion,
especially on the viscous axisymmetric flow, are the
significant steps toward understanding various astronomical
sources that are believed to be powered by the 
accretion onto black holes.
Although the theory is far from complete and we may have to wait
many years to understand the real three-dimensional hydrodynamics
and radiative transfer, it already is rather successful in
explaining most X-ray sources. In addition, the incomplete part
of our understanding could be the key to the remaining mysteries
since the most attractive current quasi spherical solutions
are likely to be unstable to the formation of (unsteady) jets.

ACKNOWLEDGMENTS

We thank Xingming Chen and Insu Yi for useful discussions,
especially Xingming Chen for kindly providing the data for Figure 3.
This work is partly supported by the 
Professor Dispatch Program of Korea Research Foundation 
and NSF grant AST 94-24416.

REFERENCES

Abramowicz, M. A., X. Chen, S. Kato, J.-P. Lasota, and O. Regev,
  {\it Astrophys. J.}, {\bf 438}, L37 (1995). \\
Abramowicz, M. A., B. Czerny, J. P. Lasota, and E. Szuszkiewicz,
  {\it Astrophys. J.}, {\bf 332}, 646 (1988). \\
Begelman, M. C.,
  {\it Mon. Not. R. Ast. Soc.}, {\bf 184}, 53 (1978). \\
Bisnovatyi-Kogan, G. S., and S. I. Blinnikov,
  {\it Mon. Not. R. Ast. Soc.}, {\bf 191}, 711 (1980). \\
Blondin, J. M.,
  {\it Astrophys. J.}, {\bf 308}, 755 (1986). \\
Bondi, H., 
  {\it Mon. Not. R. Ast. Soc.}, {\bf 112}, 195 (1952). \\
Chang, K. M., and J. P. Ostriker, 
  {\it Astrophys. J.}, {\bf 288}, 428 (1985). \\
Chakrabarti, S. K.,
  {\it Phys. Rep.}, {\bf 266}, 229 (1996). \\
Chakrabarti, S. K., and D. Molteni,
  {\it Astrophys. J.}, {\bf 417}, 671 (1993). \\
Chen, X., M. A. Abramowicz, and J.-P. Lasota,
  {\it preprint}, astro-ph/9607020 (1996). \\
Chen, X., M. A. Abramowicz, and J.-P. Lasota,
  R. Narayan, and I. Yi,
  {\it Astrophys. J.}, {\bf 443}, L61 (1995). \\
Chen, X., and R.E. Taam,
  {\it Astrophys. J. in press} (1996). \\
Cowie, L. L., J. P. Ostriker, and A. A. Stark,
  {\it Astrophys. J.}, {\bf 226}, 1041 (1978). \\
Flammang, R. A.,
  {\it Mon. Not. R. Ast. Soc.}, {\bf 199}, 833 (1982). \\
Flammang, R. A.,
  {\it Mon. Not. R. Ast. Soc.}, {\bf 206}, 589 (1984). \\
Frank, J., A. R. King, and D. J. Raine,
  {\it Accretion Power in Astrophysics},
  Cambridge U. Press, Cambridge (1985). \\
Grindlay, J. E.,
  {\it Astrophys. J.}, {\bf 221}, 234 (1978). \\
Krolik, J. H., and R. A. London,
  {\it Astrophys. J.}, {\bf 167}, 18 (1983). \\
Kusunose, M., and S. Mineshige,
  {\it Astrophys. J.}, {\bf 440}, 100 (1995). \\
Lasota, J.-P., M. A. Abramowicz, X. Chen, J. Krolik,
  R. Narayan, and I. Yi,
  {\it Astrophys. J.}, {\bf 462}, 142, (1996). \\
Lightman, A. P., and D. M. Eardley,
  {\it Astrophys. J.}, {\bf 187}, L1 (1974). \\
Loeb, A., and A. Laor, 
  {\it Astrophys. J.}, {\bf 384}, 115 (1992). \\
Maraschi, L., R. Roasio, and A. Treves,
  {\it Astrophys. J.}, {\bf 253}, 312 (1982). \\
Melia, F.,
  {\it Astrophys. J.}, {\bf 387}, L25 (1992). \\
Melia, F.,
  {\it Astrophys. J.}, {\bf 426}, 577 (1994). \\
M\'esz\'aros, P.,
  {\it Astr. Astrophys.}, {\bf 44}, 59 (1975). \\
Michel, F. C., 
  {\it Astrophys. Space Sci.}, {\bf 112}, 195 (1972). \\
Mihalas, D., and B. W. Mihalas,
  {\it Foundations of Radiation Hydrodynamics},
  Oxford University Press, Oxford (1984). \\
Molteni, D., G. Lanzafame, and S. K. Chakrabarti,
  {\it Astrophys. J.}, {\bf 425}, 161 (1994). \\
Molteni, D., D. Ryu, and S. K. Chakrabarti,
  {\it preprint}, astro-ph/9605116 (1996). \\
Narayan, R., S. Kato, and F. Honma,
  {\it preprint}, astro-ph/9607019 (1996b). \\
Narayan, R., J. E. McClintock, and I. Yi,
  {\it Astrophys. J.}, {\bf 457}, 821 (1996a). \\
Narayan, R., and I. Yi, 
  {\it Astrophys. J.}, {\bf 428}, L13 (1994). \\
Narayan, R., and I. Yi, 
  {\it Astrophys. J.}, {\bf 444}, 231 (1995a). \\
Narayan, R., and I. Yi, 
  {\it Astrophys. J.}, {\bf 452}, 710 (1995b). \\
Narayan, R., I. Yi, and R. Mahadavan,
  {\it Nature}, {\bf 374}, 623 (1995) \\
Nobili, L., R. Turolla, and L. Zampieri, 
  {\it Astrophys. J.}, {\bf 383}, 250 (1991). \\
Novikov, I. D., and K. S. Thorne,
  in {\it Black Holes}, ed. C. DeWitt and B. DeWitt,
  Gordon and Breach, New York (1973). \\
Ostriker, J. P., R. McCray, R. Weaver, and A. Yahil,
  {\it Astrophys. J.}, {\bf 208}, L61 (1976). \\
Park, M.-G., 
  {\it Astrophys. J.}, {\bf 354}, 64 (1990a.) \\
Park, M.-G., 
  {\it Astrophys. J.}, {\bf 354}, 83 (1990b). \\
Park, M.-G.,
  {\it Astr. Astrophys.}, {\bf 274}, 642 (1993). \\
Park, M.-G., and G. S. Miller,
  {\it Astrophys. J.}, {\bf 371}, 708 (1991). \\
Park, M.-G., and J. P. Ostriker,
  {\it Astrophys. J.}, {\bf 347}, 679 (1989). \\
Piran, T.,
  {\it Astrophys. J.}, {\bf 221}, 652 (1978). \\
Pringle, J. E.,
  {\it Mon. Not. R. Ast. Soc.}, {\bf 177}, 65 (1976). \\
Pringle, J. E.,
  {\it Ann. Rev. Astron. Astrophys.}, {\bf 19}, 137 (1981). \\
Pringle, J. E., and M. J. Rees,
  {\it Astr. Astrophys.}, {\bf 21}, 1 (1972). \\
Ryu, D., G. L. Brown, J. P. Ostriker, and A. Loeb,
  {\it Astrophys. J.}, {\bf 452}, 364 (1995). \\
Shakura, N. I., and R. A. Sunyaev,
  {\it Astr. Astrophys.}, {\bf 24}, 337 (1973). \\
Shakura, N. I., and R. A. Sunyaev,
  {\it Mon. Not. R. Ast. Soc.}, {\bf 175}, 613 (1976). \\
Shapiro, S. L., 
  {\it Astrophys. J.}, {\bf 180}, 531 (1973). \\
Shapiro, S. L., A. P. Lightman, and D. N. Eardley,
  {\it Astrophys. J.}, {\bf 204}, 187 (1976). \\
Soffel, M. H.,
  {\it Astr. Astrophys.}, {\bf 116}, 111 (1982). \\
Stellingwerf, R. F.,
  {\it Astrophys. J.}, {\bf 260}, 768 (1982). \\
Stellingwerf, R. F., and J. Buff,
  {\it Astrophys. J.}, {\bf 260}, 755 (1982). \\
Szuszkiewicz, E., M. A. Malkan, and M. A. Abramowicz,
  {\it Astrophys. J.}, {\bf 458}, 474 (1996).
Treves, A., L. Maraschi, and M. Abramowicz,
  {\it Pub. Ast. Soc. Pac.}, {\bf 100}, 427 (1988). \\
Wandel, A., A. Yahil, and M. Milgrom,
  {\it Astrophys. J.}, {\bf 282}, 53 (1984). \\
Zampieri, L., J. C. Miller, and R. Turolla,
  {\it preprint}, astro-ph/9607030 (1996). \\

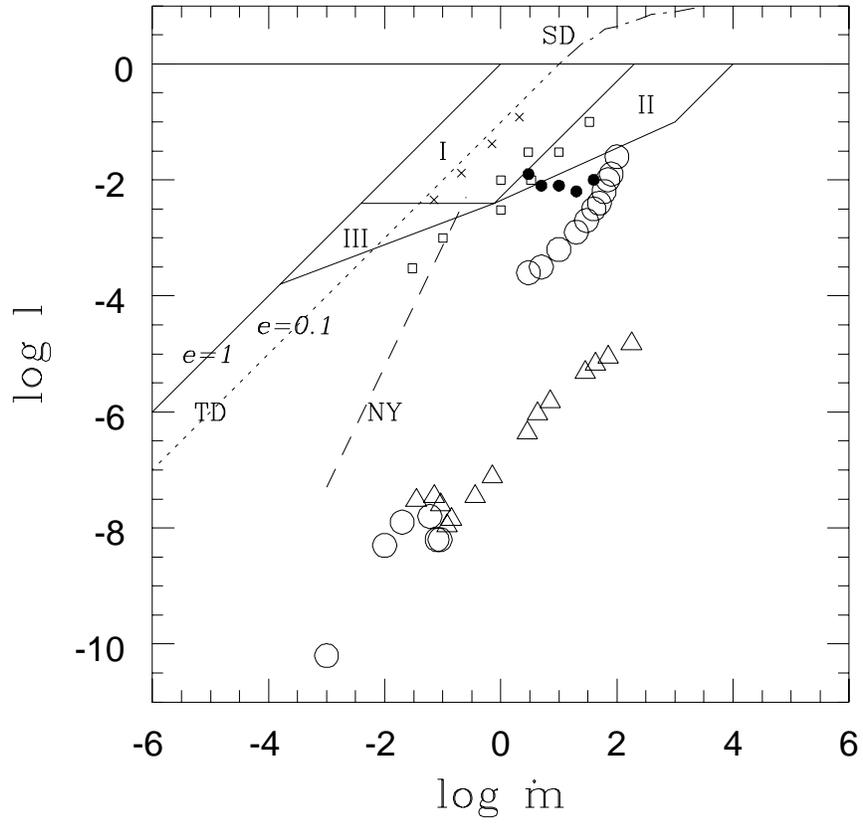
\bye